\documentclass[%
superscriptaddress,
 preprint,
 amsmath,amssymb,
 aps,
 jcp,
 letterpaper,
]{revtex4-2}

\usepackage{graphicx}
\usepackage{dcolumn}
\usepackage{bm}

\newcolumntype{.}{D{.}{.}{8}}

\usepackage[utf8]{inputenc}
\usepackage{physics}
\usepackage{amsmath, amssymb}
\usepackage{tabularx,booktabs,array,dcolumn}
\usepackage{setspace}
\usepackage{vmargin}
\usepackage{xcolor}
\usepackage{graphicx,psfrag,subfigure}

\usepackage{float}
\usepackage[all]{xy}

\usepackage{bm}
\usepackage{mathtools}
\usepackage{physics}

\newcommand{\mx}[1]{\mathbf{#1}}

\newcommand{\cm}{cm$^{-1}$}

\newcommand{\Ag}{A$_\text{g}$}
\newcommand{\Au}{A$_\text{u}$}
\newcommand{\Bg}{B$_\text{g}$}
\newcommand{\Bu}{B$_\text{u}$}

\newcommand{\Ctwoh}{$C_{2\text{h}}$}
\newcommand{\som}{Supplementary Information}

\newcommand{\eq}{\mathrm{eq}}
\newcommand{\inter}{\mathcal{I}}
\newcommand{\inttor}{$\mathcal{I}$t}
\newcommand{\intben}{$\mathcal{I}$b}
\newcommand{\inttorben}{$\mathcal{I}$tb}

\usepackage[unicode]{hyperref}
\hypersetup{
   unicode=true,          
   plainpages=false,
   colorlinks=true,       
   citecolor=blue,        
}

\begin{document}

\title{%
Fingerprint region of the formic acid dimer: 
variational vibrational computations in curvilinear coordinates
}

\author{Alberto Mart\'in Santa Dar\'ia}

\author{Gustavo Avila}

\author{Edit M\'atyus}
\email{edit.matyus@ttk.elte.hu}

\affiliation{Institute of Chemistry, 
ELTE E\"otv\"os Lor\'and University, 
P\'azm\'any P\'eter s\'et\'any 1/A, 
1117 Budapest, Hungary}

\date{\today}
\begin{abstract}
  \noindent Curvilinear kinetic energy models are developed 
  for variational nuclear motion computations
  including the inter- and the 
  low-frequency intra-molecular degrees of freedom of the formic acid dimer.
  The coupling of the inter- and intra-molecular modes is
  studied by solving the vibrational Schrödinger equation for a series of vibrational models, from two 
  up to ten active vibrational degrees of freedom by selecting various combinations of active modes and constrained coordinate values.
  Vibrational states, nodal assignment, and infrared vibrational intensity information is 
  computed using the the full-dimensional 
  potential energy surface (PES) and electric dipole moment surface
  developed by Qu and Bowman [Phys. Chem. Chem. Phys. 18, 24835 (2016); J. Chem. Phys. 148, 241713 (2018)].
Good results are obtained for several fundamental and combination bands in comparison with with jet-cooled vibrational spectroscopy experiments,  but the description of the $\nu_8$ and $\nu_9$ fundamental vibrations, which are
close in energy and have the same symmetry, appears to be problematic. 
For further progress in comparison with experiment, 
the potential energy surface, and in particular, 
its multi-dimensional couplings representation, 
requires further improvement.
\end{abstract}

\maketitle

%
%
\clearpage
\section{Introduction}
\noindent In a recent article, Nejad and Suhm reviewed the spectroscopy of the formic acid dimer (FAD) \cite{nejad2020concerted}
with a focus on the intermolecular vibrational range. 
The formic acid is the simplest carboxylic acid and its dimer is a prototype for a cyclically arranged pair of hydrogen bonds. 
The formic acid dimer has been studied in spectroscopy experiments for decades, but the rotational and temperature effects made the detection of the precise vibrational band positions challenging \cite{nejad2020concerted}. 

Coupling infrared and Raman spectrometers to jet-cooled helium beams seeded with a small amount of molecules in the gas phase 
made it possible to gain precise information on the vibrational dynamics of molecular complexes and clusters \cite{HaScSu99,LiWeSu04,LaZiSu07,SuKo13}. 
In particular, all intermolecular vibrational fundamentals and several overtone and combination bands of FAD \cite{HeGeHeHu00,georges2004jet,ZiSu07,XuSu09,KoLaDoNoSu12} have been assigned over the past 16 years with an experimental uncertainty on the order of 1~\cm.
The amount and the quality of the experimental data 
call for detailed and high-level quantum dynamics computations \cite{nejad2020concerted}.   

A detailed quantum dynamics computation relies on advanced methodology from
three areas of theoretical chemistry: 
a) electronic structure methodology that provides good approximate solutions to the electronic Schrödinger equation at a series of nuclear configurations over the coordinate range relevant for the nuclear motion; 
b) high-dimensional fitting or interpolation methods that build a potential energy surface (function) from the electronic energies available at points;
and 
c) rovibrational methodology that provides solution to the (ro)vibrational Schrödinger equation. 
A full-dimensional \emph{ab initio} potential energy surface for the formic acid dimer
is already available in the literature (parts a \& b) developed by Qu and Bowman (QB16-PES) \cite{QuBo16}, and this allows us to focus on the solution of the vibrational Schr\"odinger equation. 

Regarding the QB16-PES, it was obtained
as a least-squares fit of a permutationally invariant potential energy function for the 10-atomic formic acid dimer to 13475 CCCD(T)-F12a/haTZ electronic energies.
Qu and Bowman used a maximum of fourth order polynomials in the fitting and they report an 11~\cm\ `energy-weighted' root-mean-squared deviation and an absolute error
of  about 14~\cm\ for their points below 4400~\cm.
The global minimum structure of the PES  has \Ctwoh\ point-group symmetry
and the corresponding harmonic frequencies are in reasonable agreement with earlier theoretical and experimental work.

Qu and Bowman \cite{QuBo19} performed full(24)-dimensional vibrational configuration interaction (VCI) computations with the MULTIMODE computer program using
the normal-coordinate representation of the kinetic energy operator (KEO) and 
a 4-mode-representation of QB16-PES. 
They write about the low-frequency vibrational energies (below 1000~\cm) obtained in
the VCI computations that the energies are `slightly' up-shifted most likely due to the
use of rectilinear normal coordinates, which are usually not well suited for
describing floppy degrees of freedom, and possible deficiencies in the fitted PES. 
Otherwise, they estimate the VCI energies to be converged within about 10~\cm\ \cite{QuBo19}.

In the present work, we focus on the low-frequency range and study the role of the coordinate choice for this specific system. Is it necessary to use curvilinear coordinates to describe 
well the fingerprint range or normal coordinates are also appropriate?

In the next section (Sec.~\ref{sec:theocomp}), we define curvilinear internal coordinate vibrational models for FAD with focusing on the intermolecular dynamics.
Then (Sec.~\ref{sec:numres}), the computed vibrational states are analysed and compared with
the experimental data. 
Section~\ref{sec:assess} is about an assessment of the kinetic and the potential energy representations using internal and normal coordinates. 
The article ends (Sec.~\ref{sec:conc}) with a summary of the results, conclusions, and outlook
for possible future work.

%
%
\clearpage
\section{Theoretical and computational details \label{sec:theocomp}}
\noindent 
In the present work, the quantum dynamical computations were carried out using the GENIUSH \cite{MaCzCs09,FaMaCs11} computer program. The general rovibrational Hamiltonian implemented 
\cite{MeGu69,MeJMS79,Lu00,Lu03,LaNa02,YuThJe07} in this program is 
  \begin{equation}
   \label{eq:hamil}
   \begin{split}
    & \hat{H} = \frac{1}{2} \sum_{k=1}^D \sum_{l=1}^D \tilde{\text{g}}^{-1/4} \hat{p}_k G_{kl}\tilde{\text{g}}^{1/2}  \hat{p}_l\tilde{\text{g}}^{-1/4} \\
    & + \frac{1}{2} \sum_{k=1}^D \sum_{a=1}^3(\tilde{\text{g}}^{-1/4} \hat{p}_k G_{k,D+a}\tilde{\text{g}}^{1/4} + \tilde{\text{g}}^{1/4} G_{k,D+a}\hat{p}_k \tilde{\text{g}}^{-1/4})\hat{J}_a \\
    & + \frac{1}{2} \sum_{a=1}^3 G_{D+a,D+a}\hat{J}_a^2\\
    & + \frac{1}{2} \sum_{a=1}^3 \sum_{b>a}^3 G_{D+a,D+b}[\hat{J}_a,\hat{J}_b]_+ + \hat{V}
    \end{split}
   \end{equation}
where $\hat{J}_a$($a=1(x),2(y),3(z)$) are the body-fixed total angular momentum operators and $\hat{p}_k=-i\partial/ \partial q_k$ with the $q_k\ (k=1,2,\ldots,D)$ internal coordinates. The $G_{kl}=(\textbf{g}^{-1})_{kl}$ coefficients and $\tilde{\text{g}}=\text{det}(\textbf{g})$ are determined from the \textbf{g} matrix, defined as follows,
\begin{equation}
  g_{kl}= \sum_{i=1}^N m_i \textbf{t}^\text{T}_{ik} \textbf{t}_{il}; 
   \quad\quad\quad k,l = 1,2,...,D+3
  \label{eq:gmxt}
\end{equation}
where
\begin{equation}
\textbf{t}_{ik} = \frac{\partial \textbf{r}_i}{\partial q_k}; 
   \quad\quad\quad k,l = 1,2,...,D
   \label{eq:tvib}
\end{equation}
\begin{equation}
\textbf{t}_{i,D+a} = \textbf{e}_a \times \textbf{r}_i; 
   \quad\quad\quad a = 1(x),2(y),3(z)
   \label{eq:trot}   
\end{equation}
and $\textbf{r}_i$ are the body-fixed Cartesian coordinates for the $i$-th atom and $\textbf{e}_a$ represent the body-fixed unit vectors. 

$D\leq 3N-6$ is the number of the active vibrational dimensions in the system. If $D<3N-6$, then this definition of the kinetic energy operator (KEO) corresponds to imposing rigorous geometrical constraints for the fixed part of the system, and the results depend only on the constrained geometry, but they are independent on the actual coordinate representation of the constrained moiety \cite{MaCzCs09}. 
This procedure is sometimes referred to as `reduction in the $\textbf{g}$ matrix' that can be contrasted with
`the reduction in the $\textbf{G}$ matrix' (that is also common and) for which the results would depend on the actual coordinate representation
of the constrained fragments.

Regarding the present system,
the formic acid dimer has $N=10$ atoms and $3N-6=24$ vibrational degrees of freedom. 
A fully coupled, variational computation is currently out of reach for such a high number of degrees of freedom, except for a highly efficient normal-coordinate based representation that have been performed for FAD \cite{QuBo16,QuBo18jcpl,QuBo18high,QuBo18fd} and other systems of similar size \cite{ThoCarr17,ThoCarr18}. 
Solution methods for high-dimensional semi-rigid systems are in a far more advanced stage 
than methods for solving systems with (more than 1-2) floppy degrees of freedom.
Several floppy degrees of freedom in a system are typically strongly coupled
and in this case one has to rely on a direct product representation for which the computational cost scales exponentially with the number of the degrees of freedom. Efficient computational schemes (grid and basis reduction, contraction) exploit the weak coupling of modes or groups of modes. There are successful efforts in the community \cite{Wang:18} for solving high(er)-dimensional systems with several floppy modes. 
In a series of recent work \cite{AvMa19,AvMa19b,AvPaCzMa20}, we developed and used a computational scheme in which the floppy part is fully coupled and we exploit efficient grid and basis truncation schemes for the semi-rigid part. As a result, the cost of the computation scales exponentially with the number of floppy and polynomially with the number of semi-rigid degrees of freedom.

This work is about an exploratory, first application of the curvilinear methodology for FAD, a series of reduced-dimensional curvilinear internal coordinate models are defined and we solve the corresponding vibrational Schr\"odinger equation. 
For a specific coordinate choice, the GENIUSH program automatically computes the KEO coefficients over a grid, and uses the matrix representation of the Hamiltonian (Eq.~\ref{eq:hamil}) constructed with discrete variable representation (DVR) \cite{00LiCa} for the active vibrational degrees of freedom. The lowest-energy eigenvalues and corresponding eigenvectors are computed 
using an iterative Lanczos algorithm~\cite{lanczos,MaSiCs09}. 

\subsection{Curvilinear internal coordinates and body-fixed frame\label{sec:coord}}
The GENIUSH program uses the t-vector representation to construct the $\mx{g}$, and then the $\mx{G}$ matrices appearing in the KEO, 
Eqs.~(\ref{eq:gmxt})--(\ref{eq:trot}). 
To compute the vibrational t-vectors, Eq.~(\ref{eq:tvib}), it is necessary to define 
the Cartesian coordinates in the body-fixed frame, $\mx{r}_i$, with respect to the internal coordinates, $q_k$. Based on this definition (added to the program as a subroutine), the coordinate derivatives, the KEO coefficients, and the Hamiltonian matrix terms are constructed in an automated fashion. 
Any coordinates can be set to constrained, to a fixed value provided by the user, or active, for which an appropriate coordinate range and DVR must be defined.

First, we define the monomer structures using a $Z$-matrix-type notation. These coordinates belong to the intramolecular modes that are set as active or fixed simultaneously in both monomers to respect the compositional symmetry of the system.
Next, we define the coordinates that describe the relative position and orientation of the two monomers similarly to Refs.~\cite{SaCsAlWaMa16,SaCsMa17,dimers}.

The monomer coordinate definition is summarized in Figure~\ref{fig:moncoor}. 
The coordinate axes and the molecular plane are attached to the OCO fragment in both monomers and the following algorithm is implemented to define the KEO:
\begin{enumerate}
  \item 
  To define the intra-monomer coordinates,
the Cartesian coordinates of both monomers are defined as a function of the interatomic distances, $r_{ij}$,
angles, $\vartheta_{ijk}$, and torsion angles, $\tau_{ijkl}$ 
 according to the following expressions (see also Figure~\ref{fig:moncoor}):
\begin{equation}
\begin{split}
    \underline{\Tilde{r}}^{(I)}_{\text{C}_1}= \underline{0};
    \quad\quad \Tilde{r}^{(I)}_{\text{O}_2}= 
\begin{pmatrix}
0\\
0\\
-r_{12}
\end{pmatrix};
\quad\quad \Tilde{r}^{(I)}_{\text{O}_3}= 
\begin{pmatrix}
0\\
r_{13} \sin{(\pi+\vartheta_{213})} \\
r_{13} \cos{(\pi+\vartheta_{213})}
\end{pmatrix};
\\
\\
\Tilde{r}^{(I)}_{\text{H}_4}= 
\begin{pmatrix}
r_{14} \sin{(\pi-\vartheta_{214})} \sin{\tau_{4123}}\\
r_{14} \sin{(\pi-\vartheta_{214})} \cos{\tau_{4123}}\\
r_{14} \cos{(\pi-\vartheta_{214})}
\end{pmatrix};
\quad\quad \Tilde{r}^{(I)}_{\text{H}_5}= 
\begin{pmatrix}
r_{25} \sin{\vartheta_{125}} \sin{\tau_{5214}}\\
r_{25} \sin{\vartheta_{125}} \cos{\tau_{5214}} \\
r_{25} \cos{\vartheta_{125}}
\end{pmatrix}
\end{split}
\end{equation}
where $I=A,B$ labels the monomers. 
\begin{figure}
  \begin{center}
  \includegraphics[scale=0.5]{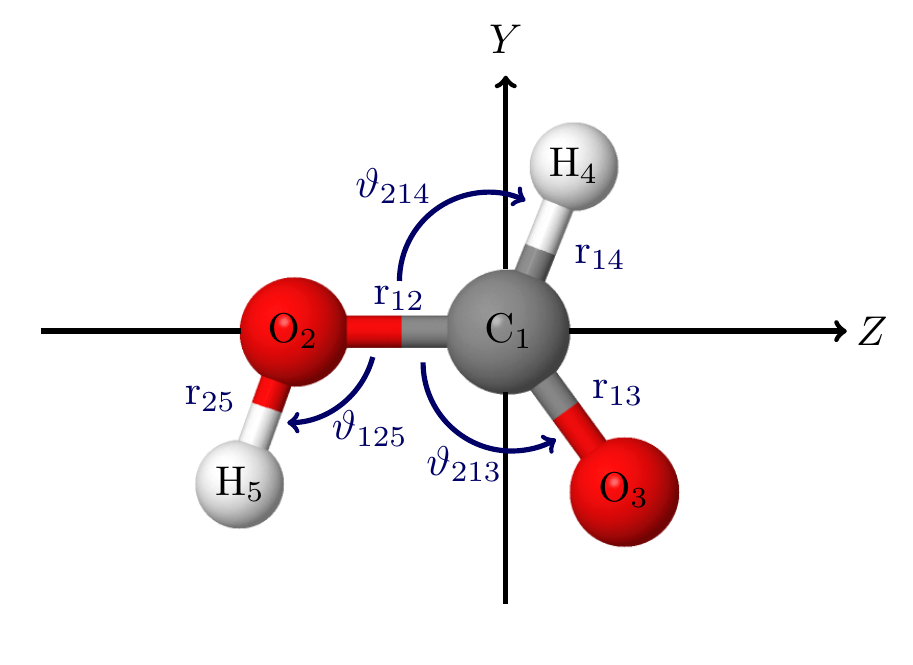}\vspace{-0.5cm}
  \raisebox{0cm}{%
  \begin{tabular}{@{}lll lll l@{}}
  \hline
  \hline\\[-0.4cm]
  C$_1^{(I)}$	& & & & & & \\
  O$_2^{(I)}$	& C$_1^{(I)}$ & $r_{12}^{(I)}$ [1.31156] & & & \\
  O$_3^{(I)}$	& C$_1^{(I)}$ & $r_{13}^{(I)}$ [1.21745] & O$_2^{(I)}$ & $\vartheta_{213}^{(I)}$ [126.14500] & & \\
  H$_4^{(I)}$	& C$_1^{(I)}$ & $r_{14}^{(I)}$ [1.09289] & O$_2^{(I)}$ & $\vartheta_{214}^{(I)}$ [111.80663] & O$_3^{(I)}$ & $\tau_{4123}^{(I)}$ [0] \\
  H$_5^{(I)}$	& O$_2^{(I)}$ & $r_{25}^{(I)}$ [0.99274] & C$_1^{(I)}$ & $\vartheta_{125}^{(I)}$ [109.72882] & H$_4^{(I)}$ & $\tau_{5214}^{(I)}$ [180] \\
  \hline\hline
  \end{tabular}
  }
  \end{center}
  \caption{
    Internal coordinate definition of the formic acid monomer.
    The same internal coordinate definition is used for both monomers, $I=A$ and $B$, 
    within the dimer.
    For the global minimum  of the dimer, 
    the values of the monomer coordinates, in \AA\ for the distances and in degree for the angles, are shown in brackets.
  }
  \label{fig:moncoor}
\end{figure}
\item
  Shift to the monomer center of mass (CM) for each monomer:
\begin{equation}
    \underline{r}^{(I)}_j := \underline{\Tilde{r}}^{(I)}_j - \underline{r}^{(I)}_{\text{CM}} ; \quad j=1,...,5
\end{equation}
Up to this point, both monomers have identical positions.   
\item
To define the inter-molecular coordinates, we rotate both monomers from their original orientation using the rotation matrices $\underline{\underline{O}}_1(0,\theta,\phi)$ and $\underline{\underline{O}}_2(\alpha,\beta,\gamma)$ parameterized with five Euler angles, 
$(\theta,\phi,\alpha,\beta,\gamma)$,
and we shift
monomer $B$ by $R$ in the positive direction along the $z$ axis (Figure~\ref{fig:dimer}):
\begin{align}
    \underline{r}^{(A)}_j &:=
    \underline{\underline{O_1}}(0,\theta,\phi)
    \underline{r}^{(A)}_j; \quad\quad
    \underline{{r}}^{(B)}_j :=
    \underline{\underline{O_2}}(\alpha,\beta,\gamma)
    \underline{r}^{(B)}_j + \left( \begin{array}{c} 0 \\ 0 \\ R \end{array} \right) \nonumber \\
    \quad& R\in[0,\infty), \quad \theta,\beta\in[0,\pi], \quad \phi,\alpha,\gamma\in[0,2\pi)
\end{align}
\begin{figure}
    \begin{center}
      \includegraphics[scale=0.65]{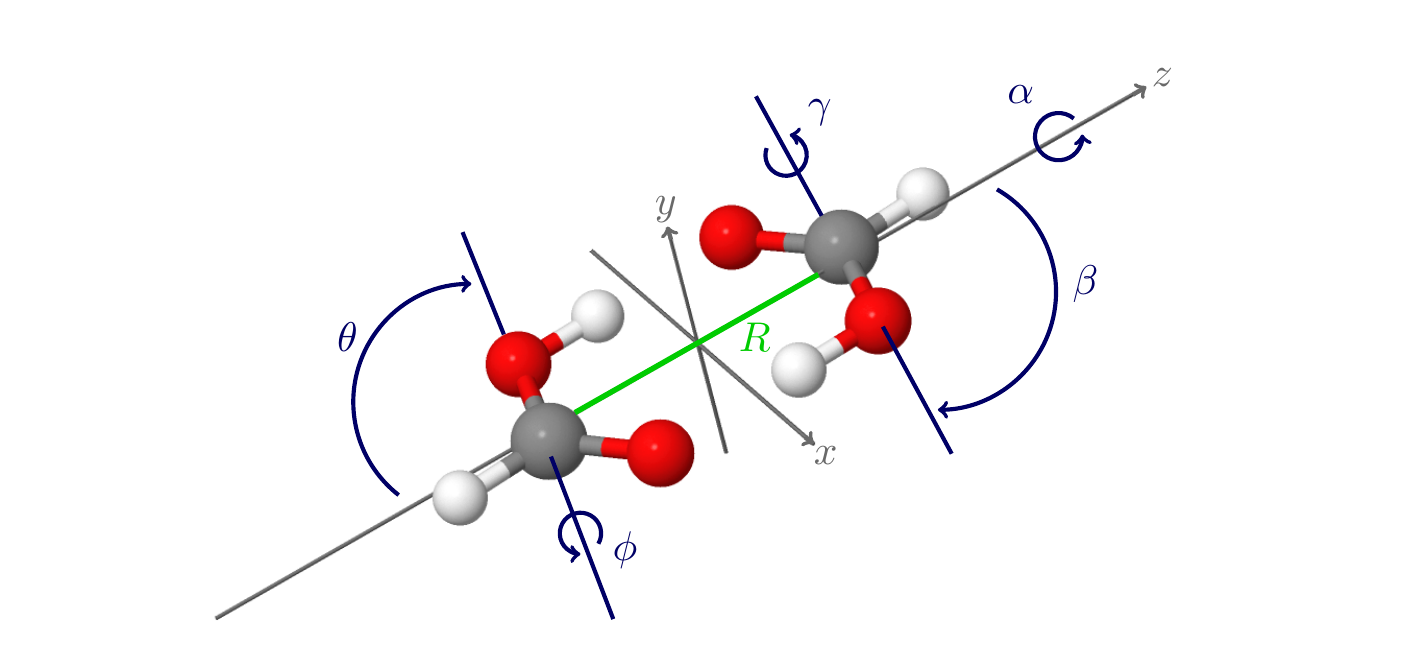}
    \end{center}      
    \caption{
      Definition of the intermolecular coordinates, $(R,\theta,\phi,\alpha,\beta,\gamma)$,
      shown for the equilibrium geometry of the dimer.
      }
    \label{fig:dimer}
\end{figure}
\item
Shift to the dimer center of mass:
\begin{equation}
    \underline{\bar{r}}^{(I)}_j 
    :=
    \underline{{r}}^{(I)}_j - \underline{r}^{(AB)}_\text{CM}
\end{equation}
\item (optional)
Change of the body fixed frame of the dimer: the overall dimer can be rotated to a new body-fixed frame, for example,
to the Eckart frame.
\end{enumerate}

The curvilinear nature of the coordinates results in singularities in the KEO. 
We have defined the coordinates so that the singularities (most importantly, 
at $\cos\theta=\pm 1$ and $\cos\beta=\pm 1$) of the KEO are possibly
far from the equilibrium structure and the dynamically important coordinate range.
For example, it would be a more natural choice to align the C--H bond of both monomers along the $z$
axis, but then the equilibrium structure of the dimer would correspond to $\theta=0$
($\cos\theta=1$) where the KEO has a singularity. For this reason, we align the C=O bond in both monomers
along the $z$ axis (Figure~\ref{fig:dimer})
and perform the $\underline{\underline{O}}_1(0,\theta,\phi)$ and $\underline{\underline{O}}_2(\alpha,\beta,\gamma)$ rotation from this initial orientation. As a result the values of the angles at the global minimum (Table~\ref{tab:models})
are far from the singularities of the KEO.

Throughout this work, we used the atomic masses $m(\text{H})=1.007825\ m_\text{u}$, $m(\text{D})=2.014000\ m_\text{u}$, $m(\text{C})=12\ m_\text{u}$, and $m(\text{O})=15.994915\ m_\text{u}$ \cite{sansonetti2005handbook}.

\subsection{Vibrational models and matrix representation}
\begin{figure}
    \includegraphics[scale=0.85,angle=0]{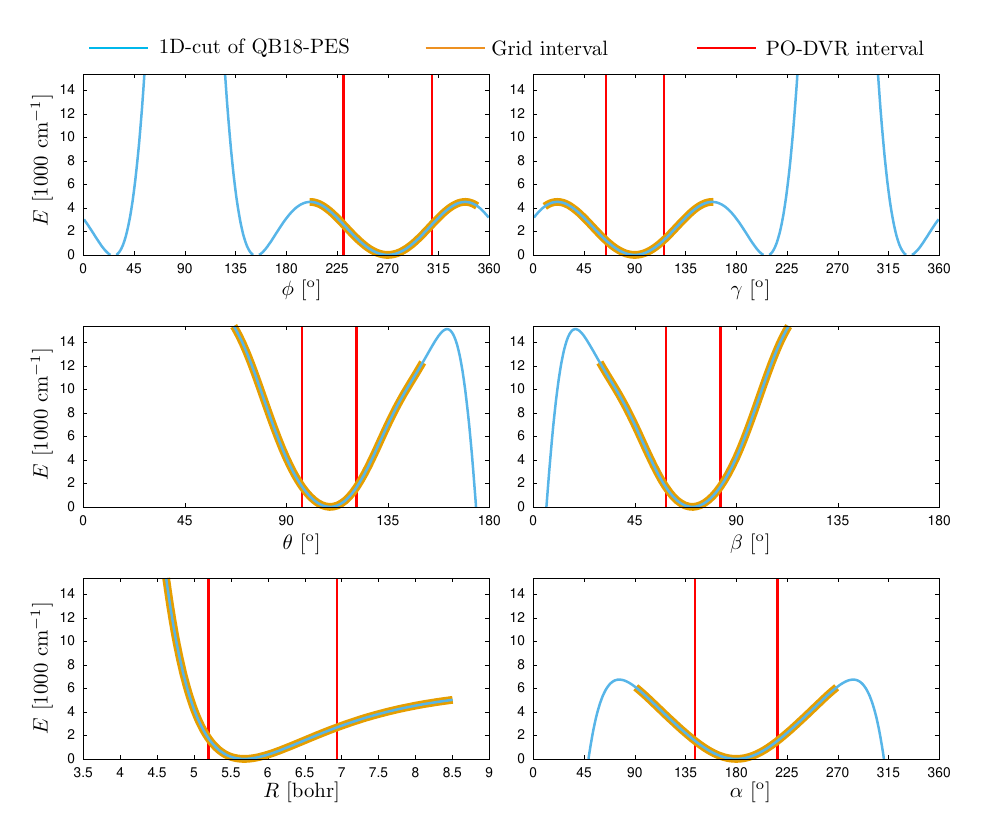}
    \caption{%
      1-dimensional cuts of the QB16-PES along the intermolecular coordinates. The intervals used in the 1-dimensional DVR computations are highlighted in orange and the PO-DVR points used in the multi-dimensional vibrational computations are located within the intervals surrounded by the red lines.
    \label{fig:1Dcut}
    }
\end{figure}

For a start, the constrained coordinates were fixed at their equilibrium value of the dimer's global minimum.
Regarding the active coordinates, their initial range was determined from inspection of 1-dimensional cuts of the QB16-PES (Figure~\ref{fig:1Dcut}). Based on these 1D cuts, we may expect that the fingerprint region can be well described.
For every coordinate, we performed a 1D vibrational computation using the 1D (unrelaxed) potential energy cut
over a broad, physically  meaningful interval (highlighted in orange in the figure) 
that is not affected by an unphysical behavior of the PES.
For this 1D model, we employed a large number of a discrete variable representation (DVR)  points \cite{00LiCa} scaled to the selected interval.
These computations were used to define potential-optimized DVRs (PO-DVR) \cite{WeCa92,EcCl92}.

This construction (and the semi-rigid properties of the system) allowed us to retain 
only a small fraction of grid points for the multi-dimensional vibrational computations.
Table~\ref{tab:models} summarizes the coordinate parameters 
(minimum value and interval), 
the DVR grid type and the number of points, 
as well as the number of PO-DVR points 
that were found to be sufficient to converge the multi-dimensional variational computations
presented in this work within 0.01~\cm. This convergence threshold is orders of magnitude 
better than the fitting error reported for the PES, but we used this threshold to
make sure that all states appearing in our energy list are true (converged) states corresponding to the vibrational model.

The reported computations were not very expensive, even the  largest  one  finished  within  1-2  days  using  24  cores. At the same time, the  largest computations required  a  large  amount  of  memory ($\sim 1$~TB), since we used a simple direct product representation and all eigenvectors for the reported states were generated simultaneously in memory.  For higher-dimensional computations, we plan to exploit the Smolyak scheme \cite{tc-gab1,tc-gab2,AvCa11b} recently implemented in GENIUSH \cite{AvMa19,AvMa19b} (and perhaps start further developments).

\begin{table}
\caption{%
  Coordinate intervals and representations. 
}
  \label{tab:models}

\begin{tabular}{@{}lccccc@{}}
\hline\\[-0.35cm] 
\hline\\[-0.35cm] 
  \raisebox{-0.20cm}{Coordinate} & 
  \raisebox{-0.20cm}{Equilibrium value} & 
  \multicolumn{3}{c}{DVR} & 
  {PO-DVR}  \\[-0.20cm]
\cline{3-5}\\[-0.35cm]
           &                   & type & \# & interval &  \#  \\
\hline\\[-0.35cm]

$R\ [\textup{\r{A}}]$ & 3.007879 & Laguerre & {300} & [2.0,4.5] & 7 \\
$\cos\theta$  & $-0.333 027$ & Legendre & {101} & [10,150] & 9 \\
$\phi\ [º]$   & 270 & Fourier & {101} & [200,350] & 9 \\
$\alpha\ [º]$ & 180 & Fourier & {101} & [90,270] & 11 \\
$\cos\beta$   & $0.333 027$ & Legendre & {101} & [30,170] & 9 \\
$\gamma\ [º]$ & 90 & Fourier & {101} & [10,160] & 9 \\
\hline\\[-0.35cm] 
$\tau^{A,B}_{5214}$(\text{HOCO})  \ [º] & 180 & Fourier & {101} & [120,250] & 9 \\
$\vartheta^{A,B}_{213}$(OCO)  \ [º] & {126.145} & {Fourier} & {101} & {[90,160]}  & 9 \\
\hline\\[-0.4cm] 
\hline\\
\end{tabular}
\end{table}

\begin{table}
\caption{%
  Monomer structural parameters: $r_\text{eq}$, equilibrium value at the global minimum (GM) and $\langle r\rangle_0^{\text{(2D)}}$ averaged structural parameter for the ground-state wave function of the 2D model in FAD.
  \label{tab:expvalues} 
}
\begin{tabular}{@{}l cccc@{}}
\hline\\[-0.35cm]
\hline\\[-0.35cm]
                       & 
                       $r^{(A,B)}_{25}$(O--H)$^\ast$ [\AA] & 
                       $r^{(A,B)}_{13}$(C=O) [\AA]  &  
                       $r^{(A,B)}_{12}$(C--O) [\AA] & 
                       $\vartheta^{(A,B)}_{213}$(OCO) [$^\mathrm{o}$]  \\
\hline\\[-0.35cm]
$r_\text{eq}$ (GM)         & 0.99274 & 1.21745 & 1.31156 & 126.14500 \\
$\langle r\rangle_0^{\text{(2D)}}$ & 1.01790 & 1.22136 & 1.31625 & 126.18119 \\
\hline\\[-0.35cm]
\hline\\[-0.35cm]
\end{tabular}
$^\ast$~Restricted to the anharmonic well, not accounting for tunneling.
\end{table}

%
%
\clearpage
\section{Analysis of the computed vibrational states \label{sec:numres}}
\noindent 
By inspecting the harmonic frequencies of FAD (listed in the \som), we may think that the inter- and intra-molecular dynamics are not perfectly separated. 
Based on the energetic ordering, 
it may be necessary to include at least 
the OCO bending, $\vartheta^{(A,B)}_{213}$, and/or
the HOCO torsional modes, $\tau^{(A,B)}_{5214}$, 
to have a correct description of the intermolecular dynamics.

For this reason, we present (Table~\ref{tab:vibcomp}) the lowest-energy 
vibrational energies computed with GENIUSH
using 
the 6D($\inter$) intermolecular, 
the 8D(\intben) intermolecular-bending, 
the 8D(\inttor) intermolecular-torsional, 
and the 10D(\inttorben) intermolecular-bending-torsional models and the full QB16-PES.
For exploratory reasons, we have constructed other reduced dimensionality models as well, 
\emph{e.g.,} by including the anharmonic O-H stretching modes, 
but they did not seem to qualitatively change the intermolecular vibrational energy pattern
(we note that the O--H tunneling effects may be accounted for using the same methodology, but it is left for future work).
We have also studied the effect of the precise value of the constrained structure by using the 
equilibrium, $r_\eq$, or vibrationally averaged, $\langle r\rangle_0$, parameters (Table~\ref{tab:expvalues}). Since we observed only minor shifts (a few\ \cm), we continue the discussion for the results obtained with using $r_\eq$ values for the constraints.

\begin{table}
\caption{Vibrational energies, in \cm, and vibrational transition intensities, in km mol$^{-1}$, with respect to the zero-point vibration 
computed with GENIUSH using internal coordinate KEOs and the QB16-PES \cite{QuBo16} and QB18-DMS \cite{QuBo18high}.
The intensity values are shown in the parentheses and only values larger than 0.05 km mol$^{-1}$ are shown, otherwise a `0' entry is printed.
\label{tab:vibcomp}}
\begin{center}
\begin{tabular}{@{}c r@{\ }l @{\ \ \ \ \ } r@{\ \ }l @{\ \ \ \ \ }  r@{\ \ }l @{\ \ \ \ \ }  r@{\ \ }l @{}}
\hline\\[-0.35cm]
\hline\\[-0.35cm]
 & 
 \multicolumn{2}{c}{6D($\inter$)} & 
 \multicolumn{2}{c}{8D(\intben)} & 
 \multicolumn{2}{c}{8D(\inttor)} & 
 \multicolumn{2}{c}{10D(\inttorben)}  \\
\#  
 & 
 \multicolumn{2}{c}{$[\mathcal{I}]$} & 
 \multicolumn{2}{c}{[$\mathcal{I}$\ \&\ $\vartheta^{A,B}_{213}$]} &
 \multicolumn{2}{c}{[$\mathcal{I}$\ \&\ $\tau^{A,B}_{4123}$]} &
 \multicolumn{2}{c}{[$\mathcal{I}$\ \&\ $\tau^{A,B}_{4123}$\ \&\ $\vartheta^{A,B}_{213}$]} \\ 
\hline\\[-0.35cm]
1	&	76	&(3.2)&	75	&(5.7) &	70	&(2.3) &	70	&(2.3)	 \\
2	&	152	&(0)  &	151	&(0.1) &	141	&(0)   &	140	&(0)	 \\
3	&	194	&(0)  &	192	&(0)   &	162	&(7.6) &	161	&(16.6)	 \\
4	&	211	&(0)  &	208	&(0)   &	191	&(0)   &	189	&(0)	 \\
5	&	227	&(0)  &	226	&(0)   &	208	&(0)   &	205	&(0)	 \\
6	&	232	&(0.2)&	230	&(1.0) &	211	&(0)   &	210	&(0)	 \\
7	&	258	&(33.0) &	256	&(32.9)&	233	&(0)   &	231	&(0.1)	 \\
8	&	271	&(0)  &	268	&(0.1) &	239	&(0)   &	238	&(0)	 \\
9	&	286	&(0)  &	283	&(0)   &	253	&(50.3)&	252	&(31.9)	 \\
10	&	302	&(0)  &	300	&(0)   &	262	&(0.1) &	259	&(0.2)	 \\
11	&	308	&(0)  &	306	&(0)   &	277	&(0)   &	274	&(0)	 \\
12	&	333	&(0)  &	330	&(0)   &	280	&(0)   &	278	&(0)	 \\
13	&	344	&(0)  &	341	&(0)   &	303	&(0)   &	301	&(0)	 \\
14	&	348	&(0)  &	345	&(0)   &	310	&(0.6) &	308	&(0.8)	 \\
15	&	361	&(0)  &	357	&(0)   &	323	&(0)   &	321	&(0)	 \\
16	&	376	&(0)  &	373	&(0)   &	325	&(0)   &	323	&(0)	 \\
17	&	384	&(0)  &	381	&(0)   &	332	&(0)   &	329	&(0)	 \\
18	&	386	&(0)  &	381	&(0)   &	347	&(0)   &	343	&(0)	 \\
\hline\\[-0.35cm]
\hline\\[-0.35cm]
\end{tabular}
\end{center}
\end{table}

Adding the HOCO torsional modes to the active degrees of freedom 
appears to be qualitatively important in the present case.
At the same time, the OCO bending modes have only a small effect on the energy pattern. We do not provide a detailed assignment
for all 6D, 8D, and 10D vibrational computations, but we list (Table~\ref{tab:vibcomp}) the computed vibrational 
energies together with the 
infrared intensities,  
\begin{align}
&A(\Tilde{\nu}_\text{f} \leftarrow \Tilde{\nu}_0) / (\text{km mol$^{-1}$}) \nonumber \\
&\quad = 2.506562213[(\Tilde{\nu}_\text{f} - \Tilde{\nu}_0) / \text{cm$^{-1}$}] 
\sum_{\alpha = x,y,z} [|\bra{\psi_f} \mu_\alpha \ket{\psi_0}|^2 / \text{Debye$^2$}] \; 
\label{eq:vibint}
\end{align}
to facilitate comparison of the various computed and experimental vibrational band origins (VBOs). For the VBOs observed by Raman spectroscopy, we obtain zero infrared intensity (numerically near zero) due to the spatial symmetry of the system.
In the jet-cooled experiments, we may assume that initially only the vibrational ground state is populated, so we compute only the transition intensity for excitation to the final state (`f') only from the ground state (`0'). The $\mu_\alpha$ ($\alpha=x,y,z$) electric dipole moment was evaluated using the QB18-DMS \cite{QuBo18high} and the  body-fixed frame defined in Section~\ref{sec:coord}. 

\begin{figure}
  \centering
  \includegraphics[scale=0.45]{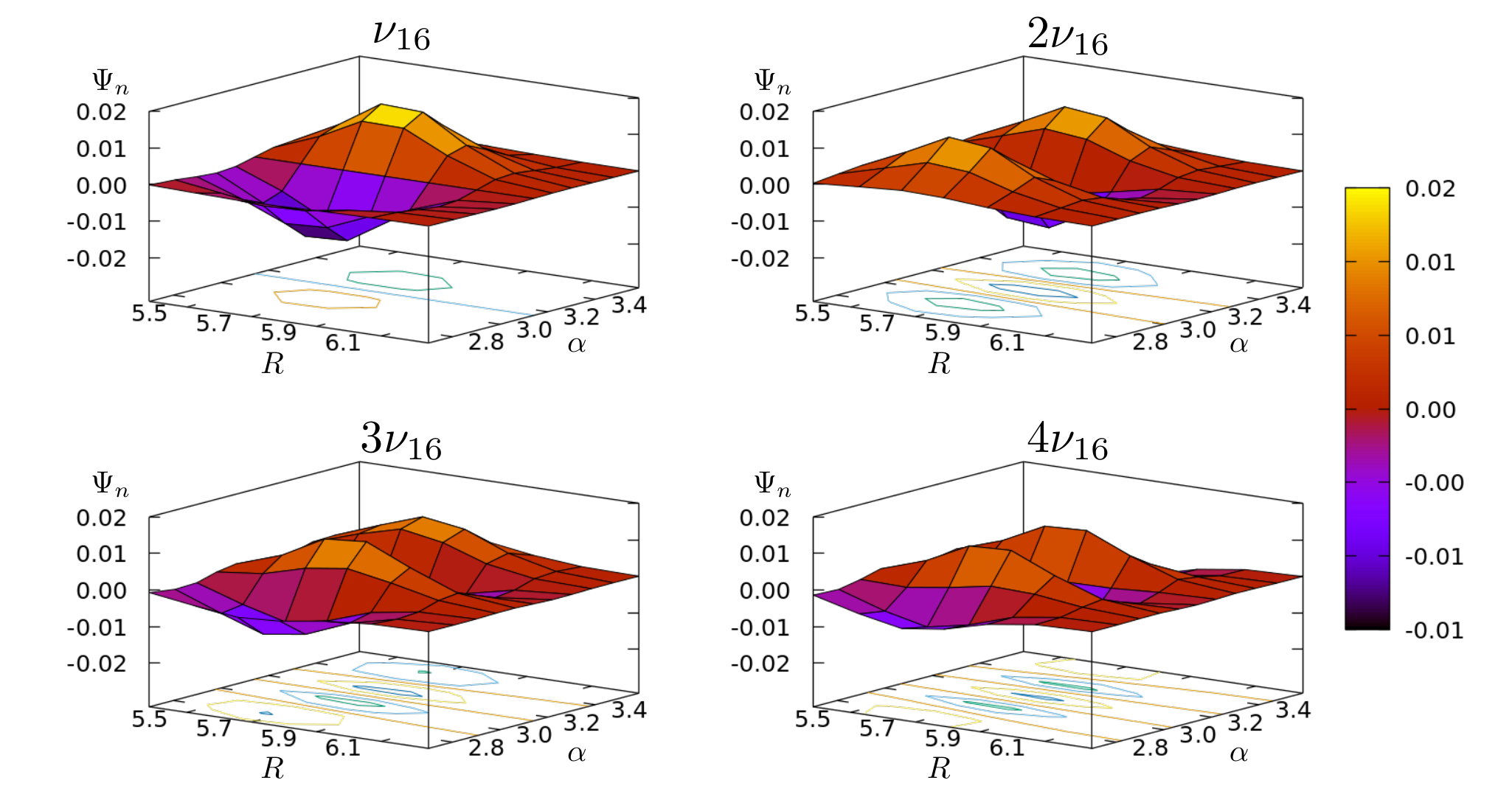} 
  \caption{%
    Example wave function plots (all quantities shown in atomic units) 
    obtained in the 8D(\inttor) computations with GENIUSH and the QB16-PES.
    \label{fig:wfpplots}
  }
\end{figure}

For the 8D(\inttor) vibrational model, we provide a detailed assignment in Table~\ref{tab:vib8Dt} for the computed vibrational states below 350~\cm\ (the first 18 vibrational states above the zero-point vibration).
The computed vibrational states were assigned based on their nodal structure.
Example wave function plots are shown in Figure~\ref{fig:wfpplots}, in which
a clean nodal structure can be observed
for the case of the fundamental, the first, the second, and the third overtone of the 
$\nu_{16}$ intermolecular twist vibration. Further wave function plots used for the assignment are deposited in the \som.

We see a less clear nodal pattern for the case of the close-lying 
states with 191 and 208~\cm\ vibrational energies, nodal features along the $R$
intermolecular stretching appear in both states. The latter state appears to have
a more pronounced stretching character, although making a decision 
between the in-plane bending ($\nu_9$) and intermolecular stretching ($\nu_8$)
would be ambiguous based on these results. The two modes belong to the same, \Ag, irreducible representation (irrep) of the \Ctwoh\ point group, and thus their mixing is allowed
(and a mixing effect has been observed already in lower-level electronic structure computations \cite{nejad2020concerted}). In comparison with the experimental results, we observe a strong 
and most likely erroneous blueshift. The erroneous behavior is indicated by the strong, 
positive `anharmonicity correction' (deviation of the variational energy from the harmonic oscillator energy) that indicates a problem in the theoretical description. 

We can identify all experimentally observed fundamental vibrations, 
overtone, and combination bands in this range, except for 
the 319~\cm\ peak tentatively assigned to the $2\nu_9$ (\Ag) band 
in Ref.~\cite{XuSu09}. This VBO may be missing from our energy list shown 
up to 347~\cm\ due to the erroneous blueshift of the $\nu_9$ fundamental vibration
in the computations. Of course, the computed energy list contains a few more combination 
and overtone bands that have not been observed in experiment yet.

So, apart from the erroneous $\nu_9/\nu_8$ system, the 8D(\inttor) 
computational results appear to be in a reasonable, although not spectacularly good,
agreement with experiment.

Switching on the OCO bending modes, resulting in the 10D(\inttorben) model,
does not change much the computed results. 
On the contrary, by freezing both the bending and the torsional degrees of freedom at
their equilibrium values, resulting in the 6D($\inter$) model, we obtain a qualitatively different
energy pattern and the problematic blue-shifts appear to be even more pronounced (Table~\ref{tab:vibcomp}). We find this result surprising, we would have expected that already the  6D($\inter$) intermolecular model is qualitatively correct and it can be further improved by adding the low-frequency intramolecular modes. To better understand the origin of this behavior, we performed a couple of test calculations using different KEOs that are reported and analyzed in Section~\ref{sec:assess}.

In summary, 8D(\inttor) appears to be the simplest vibrational model for which we get meaningful results with the present QB16 PES, but the results are far from perfect. For this reason we constrained the discussion to the first 18 states of the parent isotopologue obtained with this single model. The vibrational energies obtained for the symmetrically substituted isotopologues with the 8D(\inttor) KEO and the QB16-PES are listed in Table~\ref{tab:8dtdeut} of the Appendix without further analysis.
Later on, more states can be computed and analysed using the 8D, 10D (or perhaps even 12D) vibrational models and the most appropriate values for the constrained geometrical parameters can be determined, but we think that it is important first to clarify the origin of the erroneous blueshifts observed for some of the fundamental vibrations.

\begin{table}
  \caption{%
    Assignment and comparison with literature data of the vibrational states obtained with the 8D(\inttor) vibrational model in GENIUSH and  the QB16 PES \cite{QuBo16} and QB18 DMS \cite{QuBo18high}.
    \label{tab:vib8Dt}
  }
\begin{tabular}{@{}c rl l@{}c c r@{\ }l@{}}
\hline\\[-0.35cm]
\hline\\[-0.35cm]
\#	& 
\multicolumn{1}{c}{$\tilde\nu_\text{8D(\inttor)}$} &
\multicolumn{1}{c}{$A_\text{8D(\inttor)}$} &
\multicolumn{1}{c}{Assignment} & 
\multicolumn{1}{c}{$\tilde\nu_\text{8D(\inttor)}-\tilde\nu_\text{HO}$$^\ast$} & 
\multicolumn{1}{c}{$\tilde\nu_\text{expt}-\tilde\nu_\text{8D(\inttor)}$} & 
\multicolumn{2}{c}{$\tilde\nu_\text{expt}$}  \\
& 
\multicolumn{1}{c}{[\cm]} &	
\multicolumn{1}{c}{[km mol$^{-1}$]} &
& 
\multicolumn{1}{c}{[\cm]} &
\multicolumn{1}{c}{[\cm]} & 
\multicolumn{2}{c}{[\cm]} \\
\hline\\[-0.35cm]
1	&	70	&	(2.3)	&	$\nu_{16}$ (\Au, twist)                     &    0  & $-1$  & 69.2  & \cite{georges2004jet} \\
2	&	141	&	(0.0)	&	$2\nu_{16}$                                 &       & $-2$  & 139   & \cite{XuSu09}\\
3	&	162	&	(7.6)	&	$\nu_{15}$ (\Au, oop bend)                  &  $-5$ & $6$   & 168.5 & \cite{georges2004jet} \\
4	&	191	&	(0)	    &	$\nu_{9}$/$\nu_{8}$ (\Ag, ip bend/stre) & 21(!) &$-30$ & 161   & \cite{XuSu09} \\
5	&	208	&	(0)   	&	$\nu_{8}$/$\nu_{9}$ (\Ag, stre/ip bend) &  $-1$ &$-14$ & 194   & \cite{XuSu09} \\
6	&	211	&	(0) 	&	$3\nu_{16}$	                                &       &      &  \\
7	&	233	&	(0)	    &	$\nu_{15}+\nu_{16}$                         &       &      &  \\
8	&	239	&	(0)	    &	$\nu_{12}$ (\Bg, oop lib)          & $-15$ & 3     & 242   & \cite{XuSu09} \\
9	&	253	&	(50.3)	&	$\nu_{24}$ (\Bu, ip lib)           & $-22$ &11    & 264   & \cite{SuKo13,nejad2020concerted} \\
10	&	262	&	(0.1)	&	$\nu_{9}+\nu_{16}$                          &       &       & \\
11	&	277	&	(0)	    &	$\nu_{8}+\nu_{16}$                          &       &      & \\
12	&	280	&	(0)	    &	$4\nu_{16}$	                                &       &      & \\
13	&	303	&	(0)	    &	$\nu_{15}+2\nu_{16}$                        &       &      & \\
14	&	310	&	(0.6)	&	$\nu_{12}+\nu_{16}$                         &       &1     & 311   & \cite{KoLaDoNoSu12}\\
15	&	323	&	(0)   	&	$\nu_{24}+\nu_{16}$                         &       &      & \\
16	&	325	&	(0)	&	$2\nu_{15}$                                     &       &11    & 336   & \cite{XuSu09}\\
17	&	332	&	(0)   	&	$\nu_{9}+2\nu_{16}$                         &       &      & \\
18	&	347	&	(0)   	&	$\nu_{8}+2\nu_{16}$                         &       &      & \\
\hline\\[-0.35cm]
\hline\\[-0.35cm]
\end{tabular}
$^\ast$ The lowest harmonic frequencies, $\tilde\nu_\text{HO}$ 
corresponding to the QB16-PES are listed in column~A of Table~\ref{tab:assess}. 
\end{table}

\clearpage
\section{Assessment of the KEO and the PES representations \label{sec:assess}}
\noindent%
In order to understand better the erroneous results obtained with 
curvilinear KEOs and the QB16-PES, we performed a number of test computations (Table~\ref{tab:assess}).

First of all, we computed the normal coordinates and harmonic frequencies 
corresponding to the global minimum structure of the QB16-PES (the normal coordinate parameters are deposited as \som). 
The harmonic frequencies are listed in the A~column of Table~\ref{tab:assess}. 
Next, we have implemented the harmonic potential energy model (HO) in the GENIUSH program, 
and it can be evaluated with any types of internal coordinates and grid representations.
This offers a simple, alternative PES representation to the full QB16-PES that is used
in our DVR computations.

Regarding the KEO, we have implemented the (rectilinear) normal coordinates of the QB16-PES
as active coordinates in GENIUSH. The vibrational energies obtained for the 6D normal coordinate KEO, in which the six lowest-energy normal modes are active and all other normal coordinates fixed at the `0' value, are shown in columns B and C. The `B' column can be reproduced from combinations of the harmonic frequencies listed in column~A. Column~C contains a converged 6D normal coordinate computation on the full QB16-PES. It is interesting to note that these energies are close to the 24-dimensional normal coordinate VCI computations (converged to $\sim 10$~\cm) of Ref~\cite{QuBo19} using a 4-mode representation of the QB16-PES. 
It is also necessary to note that the `C' column is quite different from the `E' and `F' columns which correspond to 6D curvilinear KEOs and the full QB16-PES. This deviation may indicate 
(a) the deficiency of the rectilinear normal coordinates to describe floppy degrees of freedom; and/or
(b) an erratic behavior of the PES that is manifested differently for the different grid representations.

Regarding the 6D curvilinear KEOs (columns~D, E, and F), 
the results obtained with the HO-model PES (column~D) are close and slightly redshifted compared to the fully harmonic normal coordinate results (columns~A and B),
and the erroneous blueshift of the $\nu_9/\nu_8$ fundamentals is missing. 
This provides an additional check for our curvilinear KEO definition and suggests
that the unphysical deviation of the 6D--8D--10D energies (Section~\ref{sec:numres}) 
from the experimental values may originate from an unphysical behavior 
of the PES representation in the coupling of these modes.

Finally, we mention that an, in principle, numerically efficient representation  
of the intermolecular dynamics of FAD, described as a single-well system,
is provided by curvilinear normal coordinates. 
Curvilinear normal coordinates are obtained as the linear combination
of the internal coordinates defined in Sec.~\ref{sec:coord} that diagonalize
the GF matrix (the coordinate definition and the linear combination coefficients are provided in the \som).  

All in all, for further progress, it is necessary to have an improved PES. 
For the fingerprint region an improvement of the intermode coupling PES representation appears to be particularly important.

\begin{table}
\caption{%
  Assessment of the PES and the kinetic energy representations:
  vibrational energies, in \cm, measured from the zero-point vibrational energy (ZPVE).
  \label{tab:assess}  
}
\begin{tabular}{@{}c@{\ \ } c@{\ \ }c@{\ \ } c@{\ \ }c@{\ \ } c @{\ \ } c@{\ \ }c @{\ \ } c@{\ \ }c c cc@{\ \ }c@{}}
\hline\\[-0.35cm]
\hline\\[-0.35cm]
KEO: & HO$^\text{a}$ &&	
\multicolumn{2}{c}{$Q^{(6\text{D})}$$^\text{b}$ } &&
\multicolumn{2}{c}{$\xi^{(6\text{D})}$$^\text{c}$} &&
\multicolumn{1}{c}{$\mathcal{Q}^{(6\text{D})}$$^\text{d}$} & 
\multicolumn{1}{c}{$Q^{(24\text{D})}$$^\text{e}$} &
\multicolumn{1}{c}{$Q^{(24\text{D})}$$^\text{e}$} \\
\cline{4-5}\cline{7-8}\\[-0.30cm]
PES: & HO &&	HO &	full &&	HO & full && full &	4MR & 4MR \\
Label$^\text{f}$ & A	&&	B	&	C	&&	D	&	E	&&	F 	&	VSCF \cite{QuBo19}	&	VCI	\cite{QuBo19} \\
\hline\\[-0.35cm]
ZPVE          &  	&&	573	&	638	&&	562	&	682	&&	682	&	n.a.	&	n.a.	\\
$\nu_{16}$    & 70	&&	70	&	110	&&	68	&	76	&&	76	&	103	&	96	\\
$2\nu_{16}$   & 	&&	140	&	175	&&	136	&	152	&&	152	&	171	&	178	\\
$\nu_9$       & 167	&&	167	&	203	&&	163	&	194	&&	194	&	204	&	209	\\
$\nu_{15}$    & 170	&&	170	&	230	&&	166	&	211	&&	211	&	250	&	213	\\
$\nu_8$       & 209	&&	209	&	269	&&	203	&	227	&&	227	&	277	&	273	\\
$3\nu_{16}$      & 	&&	211	&	275	&&	209	&	232	&&	232	&	303	&	286	\\
$\nu_{16}+\nu_9$ & 	&&	237	&	288	&&	231	&	258	&&	258	&		&		\\
$\nu_{16}+\nu_{15}$ & 	&&	241	&	313	&&	234	&	271	&&	271	&		&		\\
$\nu_{12}$  & 254	&&	254	&	320	&&	239	&	286	&&	286	&		&		\\
$\nu_{24}$  & 275	&&	275	&	353	&&	270	&	302	&&	302	&		&		\\
& 	&&	279	&	358	&&	275	&	308	&&	308	&		&		\\
& 	&&	281	&	378	&&	277	&	333	&&	333	&		&		\\
& 	&&	308	&	387	&&	297	&	344	&&	344	&		&		\\
& 	&&	311	&	390	&&	301	&	348	&&	348	&		&		\\
\hline\\[-0.35cm]
\hline\\[-0.35cm]
\end{tabular}
\begin{flushleft}
$^\text{a}$~Harmonic oscillator model approximation corresponding to the QB16-PES.\\
$^\text{b}$~6D computation with GENIUSH using 
the QB16-PES normal coordinates in the KEO with
a harmonic oscillator model PES (HO) or with the full QB16-PES (full).
The VBOs are converged within 0.01~\cm\ with (15,15,13,13,13,13) unscaled Hermite-DVR points.
\\
$^\text{c}$~6D computation with GENIUSH using 
the curvilinear internal coordinates defined in Sec.~\ref{sec:coord} in the KEO with
a harmonic oscillator model PES (HO) or with the full QB16-PES (full).
The energies are converged with (11,11,11,11,11,11) PO-DVR points defined in Table~\ref{tab:models}.\\
$^\text{d}$~6D computation with GENIUSH using 
curvilinear normal coordinates defined in the KEO with the full QB16-PES. 
The VBOs are converged within 0.01~\cm\ using (15,15,15,15,15,15) number of unscaled Hermite-DVR points.\\
$^\text{f}$ %
Assignment corresponding to columns A and B. For the other columns analysis of the wave function would be necessary for the assignment.
\end{flushleft}
\end{table}

%
%
\clearpage
\section{Conclusion and outlook \label{sec:conc}}
\noindent %
Variational vibrational computations are reported for the 
fingerprint region of the formic acid dimer (FAD) using curvilinear 
kinetic energy operator representations and the QB16-PES. 
Besides the intermolecular coordinates, the lowest-frequency 
monomer vibrations have been included resulting in a series
of vibrational models with 6, 8, and 10 active dimensions, 
while keeping all other degrees of freedom in the system rigid.

This work was initiated by a conjecture (also mentioned by Qu and Bowman in Refs.~\cite{QuBo16,QuBo18jcpl,QuBo18fd}) that perhaps normal coordinates are not well suited and one should rather use curvilinear coordinates to describe the fingerprint region of FAD.  We hoped that by using curvilinear coordinates, it becomes possible to resolve discrepancy of earlier theory (with normal coordinates) and experiment.
During our vibrational study we obtained good results for several fundamental and combination bands in comparison with jet-cooled vibrational spectroscopy experiment, but noticed that there was a problem with the $\nu_8$ and $\nu_9$ fundamental vibrations. These vibrations have always been difficult to describe accurately (already on the harmonic level as it was pointed out by Nejad and Suhm \cite{nejad2020concerted}), because they are close in energy and have the same symmetry.
These fundamental vibrations are obtained from our variational computations with an erroneous blueshift and the series of computations with different vibrational models developed in this work suggest that for further progress in comparison with experiments, improvement of the PES is necessary.

Relying on the increasing computational resources, further development in the computational methodology, and assuming that an improved potential energy surface will become available soon, we can foresee a curvilinear treatment with more than 10 fully-coupled degrees of freedom, or studying the tunneling dynamics in the fingerprint range in the vibrational or rovibrational spectrum.
An interesting alternative future direction will be the computation of tunnelling splitting effects in the monomer stretching spectrum \cite{MaHa02,OrHa07} that was computationally studied using 
an extension of the reaction surface Hamiltonian \cite{BaSi08,BaSqSi08} and 
also with 7-dimensional curvilinear vibrational models by Luckhaus \cite{Lu06,Lu10}. Further progress in that direction also requires an improved PES representation up to a beyond the monomer stretching range.

\vspace{3cm}
\noindent \textbf{Acknowledgment}
We thank Joel Bowman and Chen Qu for sending to us their formic acid dimer PES and DMS. We thank the financial support of the Swiss National Science Foundation 
(PROMYS Grant, No.~IZ11Z0\_166525).

\clearpage
\bibliographystyle{vancouver}

\clearpage
\section{Appendix: Vibrational energies of the symmetrically deuterated isotopologues \label{sec:appendix}}

\begin{table}[htbp]
  \caption{%
    Vibrational band origins, in cm$^{-1}$, referenced to the zero-point 
    vibrational energy (ZPVE) of parent and the three symmetrically deuterated 
    isotopologues of the formic acid dimer computed with the GENIUSH program using the 
    8D($\inter$t) KEO and the QB16-PES.
  \label{tab:8dtdeut}
  }
\begin{tabular}{@{}rr rr r@{}}
\hline\\[-0.35cm]
\hline\\[-0.35cm]
\# &
\multicolumn{1}{c}{(HCOOH)$_2$}	&
\multicolumn{1}{c}{(DCOOD)$_2$}	&	
\multicolumn{1}{c}{(HCOOD)$_2$}	&	
\multicolumn{1}{c}{(DCOOH)$_2$}	\\
\hline\\[-0.35cm]
0  & 1533	&	1240	&	1280	&	1493	\\
1  & 70	    &	70	&	70	&	70	\\
2  & 141	&	135	&	140	&	138	\\
3  & 162	&	140	&	157	&	140	\\
4  & 191	&	186	&	187	&	191	\\
5  & 208	&	205	&	208	&	205	\\
6  & 211	&	207	&	210	&	208	\\
7  & 233	&	207	&	227	&	209	\\
8  & 239	&	209	&	237	&	210	\\
9  & 253	&	245	&	251	&	248	\\
10 & 262	&	256	&	257	&	261	\\
11 & 277	&	270	&	278	&	275	\\
12 & 280	&	274	&	279	&	277	\\
13 & 303	&	276	&	297	&	279	\\
14 & 310	&	277	&	307	&	279	\\
15 & 323	&	278	&	313	&	279	\\
16 & 325	&	314	&	320	&	317	\\
17 & 332	&	319	&	327	&	327	\\
18 & 347	&	326	&	341	&	331	\\
19 & 350	&	339	&	347	&	342	\\
20 & 351	&	341	&	348	&	344	\\
21 & 368	&	341	&	363	&	346	\\
22 & 372	&	343	&	366	&	347	\\
23 & 380	&	345	&	372	&	348	\\
24 & 381	&	346	&	377	&	348	\\
25 & 389	&	347	&	383	&	349	\\
26 & 392	&	371	&	389	&	380	\\
27 & 395	&	379	&	392	&	385	\\
28 & 397	&	382	&	393	&	389	\\
29 & 401	&	389	&	397	&	394	\\
\hline\\[-0.35cm]
\hline\\[-0.35cm]
\end{tabular}
\end{table}

\end{document}